\documentclass[prd,aps,nofootinbib,twocolumn]{revtex4} 

\usepackage{bm}
\usepackage{latexsym}
\usepackage{amsfonts}
\usepackage{amssymb}
\usepackage{amsmath}
\usepackage{color}

\begin{document}

\title{Ho\v{r}ava gravity with mixed derivative terms}

\author{Mattia Colombo,$^1$ A. Emir G\"umr\"uk\c{c}\"uo\u{g}lu,$^1$ and Thomas P. Sotiriou$^{1,2}$}
\affiliation{$^1$ School of Mathematical Sciences, University of Nottingham, University Park, Nottingham, NG7 2RD, UK\\
$^2$ School of Physics and Astronomy, University of Nottingham, University Park, Nottingham, NG7 2RD, UK} 

\date{\today}

\begin{abstract}

Ho\v{r}ava gravity has been constructed so as to exhibit anisotropic scaling in the ultraviolet, as this renders the theory power-counting renormalizable. However, when coupled to matter, the theory has been shown to suffer from quadratic divergences. A way to cure these divergences is to add terms with both time and space derivatives. We consider this extended version of the theory in detail. We perform a perturbative analysis that includes all modes, determine the propagators and discuss how including mixed-derivative terms affects them. We also consider the Lifshitz scalar with mixed-derivative terms as a toy model for power counting arguments and discuss the influence of such terms on renormalizability. 

\end{abstract}
\maketitle

\section{Introduction}

Einstein's General Relativity (GR) is currently in agreement with all available observational and experimental data (see e.g. \cite{Will:2005va}). However, the fact that GR is not renormalizable suggests that it is no more than a low energy effective theory. When quantum corrections are taken into account, higher derivative operators are inevitably excited \cite{Donoghue:1994dn}. The leap from effective field theory to an ultraviolet (UV) complete gravity theory is highly non-trivial. The presence of higher derivative terms in the Lagrangian does indeed improve the UV behavior of the theory through the modification that the additional spatial derivatives introduce to the propagator. However, so long as Lorentz invariance remains intact, such terms also introduce higher time derivatives, which lead to a breaking of unitarity \cite{Stelle:1976gc}. 

Based on an analogy with the Lifshitz scalars in condensed matter physics \cite{Lifshitz}, a theory of gravity which takes  time and space on a different footing was introduced by Ho\v{r}ava \cite{Horava:2009uw}. The novelty of this approach is to allow for higher spatial derivatives while restricting the kinetic part to contain no more than two time derivatives. This is achieved by breaking the isotropy in the scaling of the spatial and temporal coordinates in the UV
\begin{equation}
t \to  b^{-z} t\,,\qquad x^i \to b^{-1} x^i\,,
\label{eq:aniscal}
\end{equation}
where the critical exponent $z$ encodes the amount of scaling anisotropy. With this scaling property, the action is allowed to contain higher dimensional operators constructed only with spatial derivatives. The full 4D diffeomorphisms of GR now have to be relaxed such that the anisotropic scaling (\ref{eq:aniscal}) can be accommodated. Ho\v{r}ava's theory is defined by the ``foliation preserving diffeomorphisms'' (FDiff)
\begin{equation}
t\to \bar{t}(t)\,,
\qquad
x^i \to \bar{x}^i(t,x^i)\,.
\label{eq:fdiff}
\end{equation}
It is then constructed out of terms which are invariant under the above symmetry. Since the time coordinate is fundamentally different from the spatial ones, the Arnowitt--Deser--Misner decomposition of the 4D metric into 3D hypersurfaces of constant $t$ \cite{Arnowitt:1962hi} provides a natural description of the fundamental ingredients of the theory, in terms of the lapse function $N(t,x^i)$, the shift vector $N^i(t,x^j)$ and the spatial metric $g_{ij}(t,x^k)$.
As a result of the symmetry (\ref{eq:fdiff}), the time-kinetic part contains only quadratic terms in the extrinsic curvature $K_{ij}$, while the higher spatial derivative terms are constructed out of the 3D curvature invariants, the lapse function and their 3D covariant derivatives. For critical exponent $z=3$, the latter terms contain up to 6 spatial derivatives and constitute the minimal theory which is renormalizable at the power-counting level.

The anisotropic scaling at the level of the action is supposed to reflect the scaling of the propagator(s) or the dispersion relation(s). However, there are several reasons why this might not be the case and hence, na\"ive power counting based on this anisotropic scaling might be misleading. The power-counting arguments for Ho\v rava gravity are based on the analogy with the Lifshitz scalar (see Refs.~\cite{Visser:2009fg,Visser:2009ys} for a detailed discussion).  The latter  is a field theory of a single degree of freedom and one could  straightforwardly guess the propagator by inspection of the action. Ho\v rava gravity instead propagates a spin-2 and a spin-0 mode. In addition to those, there are also the gauge modes. It is, therefore, much more subtle to infer the behavior of the propagator by the scaling properties of the operators appearing in the action. Indeed, there exist restricted versions of the theory, such as those with {\it detailed balance} \cite{Horava:2009uw, Vernieri:2011aa, Vernieri:2012ms}, where the sixth order operators in the action do not contribute at all to the propagators of the spin-0 mode, thus compromising renormalizability. The problem can be solved by adding eighth order operators, but the main lesson from these examples is that the na\"ive anisotropic scaling that one infers from the action is not always respected by the propagators.

A further limitation of the power counting arises in determining the influence of the gauge modes on the loops. This has been demonstrated clearly in the analysis of Ref.~\cite{Pospelov:2010mp}. 
The biggest challenge for Ho\v{r}ava's theory, or any theory which violates Lorentz invariance in the gravity sector, is to suppress the Lorentz violation effects at low energy in the matter sector, where constraints are very stringent \cite{Liberati:2012jf,Liberati:2013xla}. In Ref.~\cite{Pospelov:2010mp} such a mechanism has been proposed. Lorentz violations are restricted to the gravity sector at tree level and they percolate the matter sector only though graviton loops. It is  shown that Lorentz-violating terms in the matter sector end up being suppressed by powers of  $M_\star/M_p$, where $M_\star$ is the UV scale above which the dispersion relations in the gravity sector cease to be relativistic. Hence, if $M_\star\ll M_p$, Lorentz violations in the matter sector can remain below experimental constraints.  
\footnote{Alternatively, one can introduce supersymmetry to suppress Lorentz violating operators at low energies \cite{GrootNibbelink:2004za,Jain:2005as,Xue:2010ih}, although such constructions are highly non-trivial beyond free theories \cite{Redigolo:2011bv,Pujolas:2011sk}.}

On the other hand, the analysis of Ref.~\cite{Pospelov:2010mp} also uncovered a technical naturalness problem. Gauge mode loops actually lead to quadratic divergences.\footnote{Other types of divergences, as well as a loss of unitarity were uncovered in Ref.~\cite{Kimpton:2013zb}, once matter fields are introduced.} It was shown that the problem can be solved by introducing the specific counter-term $\nabla_i K_{jk} \nabla^i K^{jk}$ that can improve the behavior of the gauge mode. This term was chosen because it does not contribute to the propagator of the spin-2 graviton. However, a thorough analysis of the effect this term, or other similar terms with mixed derivatives can have on the dynamics of the propagating modes is still pending. In addition, there is a strong ambiguity on how such terms fit in the power counting scheme. If one na\"ively tries to assign an order to them based on the scaling (\ref{eq:fdiff}) then they should be counted as eighth order operators. However, there is no reason to trust such an order assignment. Generically such terms will modify the dispersion relations of (some of) the propagating modes and could even be the leading operators with time derivatives in the UV, thus compromising anisotropic scaling altogether. The implications of having such terms in the action for renormalizability are far from obvious.

Our goal here is to shed some light into this matter. The rest of the paper is organized as follows. In the next Section we briefly review the basic ingredients of the Ho\v{r}ava gravity and we construct the action with mixed derivatives.  In Sec.~\ref{sec:perturbations} we present a full perturbation analysis of the theory and determine the propagators for all modes. This allows us to clarify the influence of the mixed-derivative terms on the propagators. Remarkably, this is the first complete perturbative analysis of (non-projectable) Ho\v{r}ava gravity, even without the mixed-derivative terms. In Sec.~\ref{sec:powercounting}, we reconsider the Lifshitz scalar as a toy model and we examine how adding  mixed-derivative term would affect power-counting renormalizability. 
We conclude with Sec.~\ref{sec:discussion} where we discuss our results.

\section{The action for Ho\v{r}ava gravity}
\label{sec:setup}

Since its introduction, Ho\v{r}ava's theory has been subject to serious scrutiny, covering a range of issues \cite{Charmousis:2009tc,Li:2009bg,Blas:2009yd,Koyama:2009hc,Papazoglou:2009fj,Henneaux:2009zb,Blas:2009ck,Kimpton:2010xi, Padilla:2010ge}, which led to the introduction of several extensions \cite{Blas:2009qj,Blas:2010hb,Horava:2010zj,Zhu:2011yu,Vernieri:2011aa}. 
A brief presentation of the various versions of the theory can be found in Refs.~\cite{Kiritsis:2009sh,Mukohyama:2010xz,Sotiriou:2010wn}.

In the rest of the paper, we will focus on the FDiff (\ref{eq:fdiff}) invariant non-projectable Ho\v{r}ava gravity in 3+1 dimensions, with critical exponent $z=3$ \cite{Horava:2009uw,Blas:2009qj}. We start by determining the most general action that is suitable for our purposes. Formally, the action we consider is
\begin{equation}
S = \frac{M_p^2}{2}\int N dt\,\sqrt{g}d^3x\left(K_{ij}K^{ij}-\lambda K^2\right)+S_V+S_{\nabla K}\,,
\label{eq:actformal}
\end{equation}
where the extrinsic curvature is defined as
\begin{equation}
K_{ij} \equiv \frac{1}{2\, N}\,\left(\dot{g}_{ij} - \nabla_i N_j - \nabla_j N_i\right)\,.
\end{equation}
The action (\ref{eq:actformal}) contains the time-derivative kinetic terms for the 3-metric $g_{ij}$, while the potential part
\begin{equation}
S_V = \int Ndt\,\sqrt{g}d^3x\left(\frac{M_p^2}{2}{\cal L}_{z=1}+{\cal L}_{z=2}+\frac{1}{M_p^2}{\cal L}_{z=3}\right)\,,
\label{eq:SV}
\end{equation}
contains up to 6 spatial derivatives and exhausts all marginal and relevant operators.
$S_{\nabla K}$ denotes all terms that are compatible with the symmetry and contain up to two time derivatives and two spatial derivatives, including the mixed-derivative term considered in Ref.~\cite{Pospelov:2010mp}. One could also add the relevant deformation ${\cal L}_{z=0} = \Lambda$ that is allowed by the FDiff symmetry and the power counting. However, since we will later focus on a Minkowski background, we will be neglecting this cosmological constant term.

The number of all possible terms in $S_V$ and $S_{\nabla K}$ is of the order $10^2$. However, we are interested in linear perturbations around flat spacetime. So, without loss of generality, we can consider only the terms that give non-trivial contributions to the propagation of linear perturbations around the Minkowski background. We expand the basic quantities as
\begin{equation}
N = 1+\delta N\,, \qquad N_i = \delta N_i\,, \qquad g_{ij} = \delta_{ij}+\delta g_{ij}\,,
\label{eq:decomp}
\end{equation}
and impose a truncation of the action at quadratic order in perturbations. The building blocks for constructing the FDiff invariant potential terms are the acceleration 3-vector (1 spatial derivative)
\begin{equation}
a_i \equiv \partial_i\log N = \partial_i \delta N +{\cal O}({\rm perturbation})^2\,,
\end{equation}
and the 3 dimensional Ricci curvature tensor (2 spatial derivatives)
\begin{eqnarray}
R_{ij} &=& - \frac{\delta^{lm}}{2}\left[\partial_l\partial_m \delta g_{ij} +\partial_i \partial_j \delta g_{lm}-2\partial_l\partial_{(i}\delta g_{j)m}\right]
\nonumber\\
&&+{\cal O}({\rm perturbation})^2\,.
\end{eqnarray}
In 3 dimensions the Weyl tensor is identically zero, so the Riemann tensor can be expressed solely in terms of the Ricci tensor and the metric.
Both $a_i$, $R_{ij}$ and their derivatives are of the order of perturbations, so any potential term which is cubic in these will be of higher order in the quadratic truncation. This observation reduces the number of possible terms considerably.

Even after restricting the terms to be quadratic in the acceleration, curvature and their derivatives, there are still several terms which are redundant at the level of the quadratic action around Minkowski. For instance, since the curvature is of the order of perturbations, we can further identify redundant terms by commuting the covariant derivatives, i.e. $\nabla_{[i} \nabla_{j]} ({\rm perturbation}) ={\cal O}({\rm perturbation})^2$. 
Moreover, performing integration by parts, some terms turn out to give the same contribution up to higher order terms in perturbative expansion, e.g.  the term $N \nabla_i R a^i$ can be written as $-N \,R \nabla_i a^i$ up to a boundary term and $ R\,a_ia^i$ (which does not contribute at the level of our quadratic truncation). Finally, making use of the contracted Bianchi identities $\nabla^j R_{ij} = \nabla_iR/2$, we find that 
the potential terms which contribute to the quadratic action are 
\begin{eqnarray}
{\cal L}_{z=1} &=& 2\alpha\,a_ia^i + \beta\,R\,,\nonumber\\
{\cal L}_{z=2} &=& \alpha_1\,R\,\nabla_i a^i+\alpha_2 \nabla_ia_j\nabla^ia^j+\beta_1 R_{ij}R^{ij}+\beta_2R^2\,,\nonumber\\
{\cal L}_{z=3} &=& \alpha_3 \nabla_i\nabla^iR\,\nabla_ja^j+\alpha_4 \nabla^2 a_i\nabla^2 a^i+\beta_3 \nabla_i R_{jk}\nabla^i R^{jk}
\nonumber\\
&&+\beta_4 \nabla_iR\nabla^iR\,,
\label{eq:act-pot}
\end{eqnarray}
where we defined $\nabla^2\equiv \nabla_i\nabla^i$.
This is the most general version of Ho\v{r}ava's theory including all terms that contribute to linear perturbations around Minkowski background. We remark that the projectable version of the theory with $N=N(t)$ can be obtained by simply taking the limit $\alpha\to\infty$ \cite{Blas:2009qj}.

We now introduce the terms we wish to focus on, which are the mixed 2-time and 2-space derivative terms. Apart from the form $(\nabla_i K_{jk})^2$ chosen in Ref.~\cite{Pospelov:2010mp}, one can also write terms of the form $(K_{ij}a_k)^2$ and $K_{ij} K_l^jR^{il}$, by appropriate contractions with the metric $g_{ij}$. However, considering the perturbed quantities (\ref{eq:decomp}), we find that
\begin{equation}
K_{ij} =\frac{1}{2}\left[\delta\dot{g}_{ij}-\partial_i\delta N_j-\partial_j \delta N_i\right] + {\cal O}({\rm perturbation})^2\,.
\end{equation}
In other words, the extrinsic curvature is also of order of perturbations; only the terms of the form $(\nabla_i K_{jk})^2$ will contribute to the quadratic action. The mixed derivative part can thus be written as
\begin{equation}
S_{\nabla K} = \int Ndt\,\sqrt{g}d^3x \nabla_iK_{jk}\nabla_lK_{mn} M^{ijklmn}\,,
\label{eq:act-dk}
\end{equation}
which consists of four independent contractions:
\begin{eqnarray}
M^{ijklmn} &\equiv &\gamma_1 g^{ij}g^{lm}g^{kn}+\gamma_2 g^{il}g^{jm}g^{kn}
+\gamma_3 g^{il}g^{jk}g^{mn}
\nonumber\\
&&+\gamma_4 g^{ij}g^{kl}g^{mn}\,.
\label{eq:dkterms}
\end{eqnarray}

The term with coefficient $\gamma_1$ corresponds to the one introduced in Ref~\cite{Pospelov:2010mp}, used to remove the quadratic divergences in the vector loops.

\section{Perturbations around Minkowski}
\label{sec:perturbations}

We now consider perturbations around flat spacetime in the non-projectable theory with mixed derivative terms, introduced in the previous Section. For a perturbative analysis of the {\it projectable} version \cite{Horava:2009uw,Sotiriou:2009gy} where the lapse function is forced to be space-independent, we refer the reader to Ref.~\cite{Sotiriou:2009bx}, and for an analysis of scalar perturbation in the non-projectable case to Refs.~\cite{Blas:2009qj, Blas:2010hb}. 

Decomposing the perturbations with respect to their transformation properties under spatial rotations, the background and perturbations are introduced as
\begin{eqnarray}
N = 1+ A\,,\qquad
N^i = (B^i+\partial^i B)\,,\qquad\qquad\qquad\nonumber\\
g_{ij}=\delta_{ij}(1+2\psi) + (\partial_i\partial_j -\frac{\delta_{ij}}{3}\partial^2)E+\partial_{(i}E_{j)}+\gamma_{ij}\,,\nonumber\\
\end{eqnarray}
where $\partial_iB^i = \partial_iE^i=\delta^{ij}\gamma_{ij} = \partial_i\gamma^{ij}=0$. We remark that since we are not working in the projectable theory, we have $A=A(t,\vec{x})$.

In the gravity sector, there are 2 tensor degrees ($\gamma_{ij}$), 4 vector degrees ($B_i$, $E_i$) and 4 scalar degrees ($A$, $B$, $E$, $\psi$), giving a total of 10 perturbations. Out of these, four will be removed by integrating out $A$, $B$ and $B_i$ (which are non-dynamical, thus entering the action without time derivatives). Furthermore, 3 degrees will be removed by exploiting the spatial transformations $x^i\to x^i+\xi^i$ (2 vectors, 1 scalar).\footnote{In the non-projectable theory, the time reparametrization invariance $t\to t+f(t)$ is not sufficient to fix any of the coordinate dependent perturbations.} In the end, we expect 3 physical degrees of freedom: 2 tensors (1 transverse traceless tensor) and 1 scalar. 

In the following, we expand perturbations into plane waves through
\begin{equation}
Q(t,\vec{x}) = \frac{1}{(2\pi)^{3/2}}\int d^3k \,Q_{\vec{k}}(t)\,e^{i\,\vec{k}\cdot\vec{x}}\,,
\end{equation}
where $Q(t,x^i)$ represents any perturbation and $Q_{\vec{k}}(t)$ is the corresponding mode function, satisfying the reality condition $Q_{-\vec{k}}=Q^\star_{\vec{k}}$. 
Thanks to the invariance of the Minkowski background under spatial rotations, the resulting quadratic action will depend only on the magnitude of the momentum $k\equiv|\vec{k}|$ and all sectors will decouple from the each other. In the remainder of the text, we omit the subscript $\vec{k}$ in the mode functions $Q_{\vec k}$.

\subsection{Tensor sector}
The action quadratic in tensor perturbations is obtained as 
\begin{eqnarray}
S_{\rm tensor}^{(2)} &=& \frac{M_p^2}{8}\int dt \,d^3k \,a^3 \left(1+2\gamma_2\kappa^2\right)
\nonumber\\&&\times\left(\vert\dot{\gamma}_{ij}\vert^2-k^2\,\frac{\beta -2\,\beta_1\kappa^2-2\,\beta_3\kappa^4}{1+2\,\gamma_2\kappa^2}\vert\gamma_{ij}\vert^2\right)\,,\nonumber\\
\end{eqnarray}
where we have defined $\kappa\equiv k/M_p$ for convenience.
Firstly, we see that only the second term of Eq.~(\ref{eq:dkterms}) contributes to the tensorial action. This is the term specifically and deliberately  omitted in the analysis of Ref.~\cite{Pospelov:2010mp}. The rest of the terms involve only divergences and traces of $K_{ij}$ and  hence, they do not contribute to the tensor sector. 
Secondly, the dispersion relation in the UV behaves as
\begin{equation}
\omega_{\rm tensor}^2= -\frac{\beta_3}{\gamma_2M_p^2}k^4 + {\cal O}(k^{2})\,,
\end{equation}
in contrast with the standard Ho\v{r}ava result with $\omega^2 \sim -\beta_3 k^6/M_p^4$.
On the other hand, tuning $\gamma_2$ to be zero reinstates the sixth order dispersion relations.

\subsection{Vector sector}
We now consider the vector sector. The quadratic action for these modes is
\begin{equation}
S_{\rm vector}^{(2)} = \frac{M_p^2}{4}\int dt\,d^3k\, k^2[1+\kappa^2(\gamma_1+2\gamma_2)]
\left\vert B^i - \frac{\dot{E}^i}{2}\right\vert^2\,.
\label{eq:vecact}
\end{equation}
In coordinate space, the equation of motion for the non-dynamical mode $B_i$ is given by
\begin{equation}
\left(1-\frac{(\gamma_1+2\gamma_2)}{M_p^2}\triangle\right)\triangle \left(B^i-\frac{\dot{E}^i}{2}\right) =0\,,
\end{equation}
where $\triangle\equiv \delta^{ij}\partial_i\partial_j$ is the the flat-space Laplace operator.
If we impose, as a boundary condition, that all perturbations and all their derivatives asymptotically vanish, then the unique solution is 
\begin{equation}
B^i = \frac{1}{2}\,\dot{E}^i\,.
\label{eq:vecB}
\end{equation}
Replacing this solution back in the action, we find that the action vanishes up to boundary terms. Hence, there are no propagating vector modes. It is clear, however, that the $\gamma_1$ and $\gamma_2$ terms modify the behavior of the vector modes by introducing extra spatial derivatives. This is exactly the feature that removed the divergences related to the vector modes in Ref.~\cite{Pospelov:2010mp}.

\subsection{Scalar sector}
The scalar action is found to be 
\begin{widetext}
\vspace{-.4cm}
\begin{eqnarray}
S_{\rm scalar}^{(2)}&=&\frac{M_p^2}{2}\int dt\,d^3k\,\Bigg\{
\left[3(1-3\lambda)+2(\gamma_1+3\gamma_2+9\gamma_3+3\gamma_4)\kappa^2\right]\left\vert \dot{\psi}+\frac{k^2}{6}\dot{E}\right\vert^2
+2\,k^2(\alpha+\alpha_2\kappa^2+\alpha_4\kappa^4) \left\vert A\right\vert^2
\nonumber\\
&&\qquad\qquad\qquad\quad
+2k^2 \left[\beta+2 (3\beta_1+8\beta_2)\kappa^2+2(3\beta_3+8\beta_4)\kappa^4\right]\left\vert \psi+\frac{k^2}{6}E\right\vert^2
\nonumber\\
&&\qquad\qquad\qquad\quad
+k^4\left[1-\lambda+2(\gamma_1+\gamma_2+\gamma_3+\gamma_4)\kappa^2\right]\left\vert B -\frac{\dot{E}}{2}\right\vert^2
\nonumber\\
&&\qquad\qquad\qquad\quad
+2\,k^2(\beta-2\alpha_1\kappa^2+2\alpha_3\kappa^4)\left[A^\star \left(\psi+\frac{k^2}{6}E\right)+{\rm c.c.}\right]
\nonumber\\
&&\qquad\qquad\qquad\quad
+k^2\left[1-3\lambda+2(\gamma_1+\gamma_2+3\gamma_3+2\gamma_4)\kappa^2\right]\left[ \left(B -\frac{\dot{E}}{2}\right)^\star\left(\dot{\psi}+\frac{k^2}{6}\dot{E}\right)+{\rm c.c}\right]\Bigg\}\,,
\label{eq:actscalar}
\end{eqnarray}
\end{widetext}
where ``c.c.'' denotes the complex conjugate of the preceding expression. Observing that the combinations $\psi+k^2\,E/6$ and $B-\dot{E}/2$ are 3D diffeomorphism invariant, the invariance of the above action is manifest.
The action now contains two non-dynamical modes that are solved by
\begin{eqnarray}
B &=& -\frac{1-3\lambda+2(\gamma_1+\gamma_2+3\gamma_3+2\gamma_4)\kappa^2}{1-\lambda+2(\gamma_1+\gamma_2+\gamma_3+\gamma_4)\kappa^2}\,\left(\frac{\dot{\psi}}{k^2}+\frac{\dot{E}}{6}\right)\nonumber\\
&&+\frac{\dot{E}}{2}
\,,\nonumber\\
A &=& -\frac{\beta-2\alpha_1\kappa^2+2\alpha_3\kappa^4}{\alpha+\alpha_2\kappa^2+\alpha_4\kappa^4}\left(\psi+\frac{k^2}{6}E\right)\,.
\end{eqnarray}
Once these solutions are inserted back into the action, the remaining terms depend on $E$ and $\psi$; more specifically, only on the gauge invariant quantity
\begin{equation}
\Psi \equiv \psi+ \frac{k^2}{6} E\,,
\end{equation}
while the remaining (pure gauge) combination drops out of the action. Thus, we arrive to
\begin{widetext}
\begin{equation}
S_{\rm scalar}^{(2)} = M_p^2 \int dt \,d^3k \left(  \frac{1-3\lambda+p_2\kappa^2+p_4\kappa^4}{1-\lambda+r_2\kappa^2}\vert \dot{\Psi}\vert^2 - M_p^2 \frac{q_2\kappa^2+q_4\kappa^4+q_6\kappa^6+q_8 \kappa^8+q_{10}\kappa^{10}}{\alpha+\alpha_2\kappa^2+\alpha_4\kappa^4}\left\vert \Psi\right\vert^2\right)\,,
\label{eq:scalaraction}
\end{equation} 
\end{widetext}
where 
\begin{eqnarray}
p_2&\equiv& 2\gamma_1(1-2\lambda)+2\gamma_2(2-3\lambda)+2(3\gamma_3+\gamma_4)\,,\nonumber\\
p_4&\equiv& 4\gamma_2(\gamma_1+\gamma_2+3\gamma_3+\gamma_4)+8\gamma_1\gamma_3-2\gamma_4^2\,,\nonumber\\
r_2&\equiv& 2(\gamma_1+\gamma_2+\gamma_3+\gamma_4)\,,\nonumber\\
q_2 &\equiv&\beta(\beta-\alpha)\,,\nonumber\\
q_4 &\equiv& -\beta(4\,\alpha_1+\alpha_2)-2\alpha(3\beta_1+8\beta_2)\,,\nonumber\\
q_6&\equiv& 4\alpha_1^2 + \beta(4\alpha_3-\alpha_4)-2\alpha(3\beta_3+8\beta_4)
\nonumber\\
&&-2\,\alpha_2(3\beta_1+8 \beta_2)\,,\nonumber\\
q_8&\equiv& -8 \alpha_1\alpha_3-2\alpha_4(3\beta_1+8\beta_2)-2\alpha_2(3\beta_3+8\beta_4)\,,\nonumber\\
q_{10} &\equiv& 4\alpha_3^2-2\,\alpha_4(3\beta_3+8\beta_4)\,.
\end{eqnarray}

Let us first recall that in the absence of the terms (\ref{eq:act-dk}), i.e.~in standard Ho\v{r}ava's theory the dispersion is $\omega^2 \propto k^6$ in the UV. In the presence of the $(\nabla K)^2$ terms (\ref{eq:act-dk}) and for generic $\gamma_i$, the coefficient of $\vert\dot{\Psi}\vert^2$ goes as $k^2$ in the UV. As a result, the dispersion relation becomes $\omega^2\propto k^4$. 
In the case of the tensor modes, a sixth order dispersion relation can be obtained by tuning only $\gamma_2$ to zero. This is still not sufficient for having $z=3$ anisotropic scaling for the scalar mode. One needs to further impose the relation $\gamma_4^2=4\gamma_1\gamma_3$ so that the $p_4$ coefficient in the kinetic term will vanish. With this tuning, the kinetic term now is a constant in the UV, giving a dispersion relation $\omega^2 \propto k^6$ despite the existence of the high order terms. 
Finally, the vector action (\ref{eq:vecact}) is only sensitive to $\gamma_1$ and $\gamma_2$ terms. Therefore, in order to simultaneously improve the quadratic UV divergences in the gauge modes {\it and} to recover sixth order dispersion relations for the propagating modes, the necessary tuning is
\begin{equation}
\gamma_2 = \gamma_4^2-4\,\gamma_1\gamma_3=0
\,,\qquad
\gamma_1 \neq 0\,.
\label{eq:tuning}
\end{equation}
For the case considered in Ref.~\cite{Pospelov:2010mp}, only the $\gamma_1$ term is non-zero and the above conditions are trivially satisfied.

We end this Section by noting that in the projectable limit $\alpha\to \infty$, the second term of Eq.(\ref{eq:actscalar}) is dominated by the $k^6$ term in the UV, while the kinetic term remains unaffected. Therefore, we conclude that the tuning  (\ref{eq:tuning}) also results in a sixth order scalar dispersion relation in the projectable version.

\section{Power counting in the presence of mixed derivative terms}
\label{sec:powercounting}

In the previous Section, we have found that in the presence of the mixed derivative term $\nabla_iK_{jk} \nabla_l K_{mn}$ the dispersion relations of the propagating degrees reduce to fourth order ones, as opposed to the sixth order in standard Ho\v{r}ava gravity. This appears to compromise power-counting renormalizability, given the fact that the latter is argued based on $z=3$ anisotropic scaling of the propagators. Our result indicates that it is actually possible to choose the coefficients of the mixed-derivative terms in such a way so as to retain sixth order dispersion relations for all modes, and still modify the UV behavior of the vector modes. So, one could potentially avoid the divergences uncovered in Ref.~\cite{Pospelov:2010mp} and still maintain $z=3$ anisotropic scaling in the UV for all modes, but this would require tuning for the coefficients of the mixed-derivative terms.

Our next step is to explore whether such tuning is indeed still necessary for power-counting renormalizability once mixed-derivative terms have been added. 
Recall that the main motivation in introducing this tuning is based on the bias that a fourth order dispersion relation is not power-counting renormalizable. However, this expectation arises from the power-counting performed in the presence of canonical kinetic terms. When mixed-derivative terms are included, the canonical kinetic term does not have to be dominant in the UV. It is therefore not at all obvious that the usual power counting argument continues to hold. 

In order to concretely discuss this issue in a simplified setting, we focus on the Lifshitz scalar in D+1 dimensions. This is anyway the basis of all power-counting arguments in Ho\v{r}ava gravity. We consider the Lagrangian
\begin{equation}
{\cal L} =\alpha \,\dot{\phi}^2 -\beta\,\dot{\phi}\triangle\dot{\phi}-\gamma \,\phi (-\triangle)^z \phi\,.
\label{eq:mixedlifshitz}
\end{equation}
Let us allow for an arbitrary anisotropic scaling
\begin{equation}
t\to b^{-m} t\,,\qquad
x^i\to b^{-1}\,x^i\,.
\label{eq:aniscalgeneral}
\end{equation}

In the standard case where $\beta=0$, renormalizability requires that $z=m=D$ \cite{Visser:2009fg, Visser:2009ys}. With these choices one can set $\alpha=\gamma=1$ without loss of generality, and the scalar field turns out to be dimensionless. It is then straightforward to argue that, if interactions of the type $g_n\phi^n$ are added, $g_n$ will have positive momentum dimensions for any $n$, a standard sign of renormalizability. Let us suppose now $\beta\neq 0$ and try to treat the corresponding term as a deformation of the standard case while retaining the same scaling dimensions. Being quadratic in both temporal and spatial derivatives, this term would (na\"ively) be an 8th order operator when $D=3$, so one arrives at a contradiction: it can hardly be considered as a simple deformation. In fact, one expects this term to be the dominant operator with time derivatives in the UV.

As we will see below, even if one considers the mixed-derivative term as a leading operator in the UV and attempts to change the scaling dimensions accordingly, ambiguities still remain. Although we find that the dimensional argument is inadequate, it demonstrates how the interpretation of the mixed term as a deformation can bring us to misleading results.

\subsection{Dimensional counting}
\label{subsec:naivecounting}

Let us repeat the power-counting arguments in a bit more detail, this time allowing for different choices of normalization and scaling. This will highlight the potential pitfalls of power-counting arguments.   As a first example, 
we consider canonical normalization for the usual kinetic term by choosing $\alpha=1$ in Eq.(\ref{eq:mixedlifshitz}). In this normalization, we have $[\beta] = [k]^{-2}$ and $[\gamma] = [k]^{-2(z-m)}$, where $[k]$ denotes the dimension of the momentum which scales as $k \to b\,k$. Moreover, we fix the units such that the operators that we expect to be dominant in the UV have the same scaling rule, imposing $[\beta] = [\gamma]$, or $m = z-1$. This allows us to rewrite the Lagrangian in the following form
\begin{equation}
{\cal L}_{1} =\dot{\phi}^2 -\frac{1}{M^2}\,\dot{\phi}\triangle\dot{\phi}-\frac{\lambda}{M^2}\,\phi (-\triangle)^z \phi\,,
\label{eq:naiveexample1}
\end{equation}
where $\lambda$ is a dimensionless constant and $M$ is some scale with dimensions of  momentum. Imposing that the action be dimensionless, we find that the momentum dimension of the scalar field is
\begin{equation}
[\phi] = [k]^{(D-m)/2}\,.
\end{equation}
This result is the same as in the canonical Lifshitz scalar case, due to the choice of normalization for the first term in (\ref{eq:naiveexample1}). The scalar field is dimensionless for $m=D$, in which case, the coefficients of non-derivative self interactions $g_n \phi^n$ have $[g_n] = [k]^{2\,D}$. However, for $m=D$ one has $z= D+1$, unlike the standard Lifshitz scalar where $z=D$. In 3+1 dimensions, this corresponds to having the usual anisotropic scaling law for the time and spatial coordinates, while the spatial derivative part of the action [the last term in eq.~(\ref{eq:naiveexample1})] is 8th order in derivatives. The mixed derivative operators would then scale as the eighth power of the momentum. 

However, the result is a by-product of the specific normalization adopted in eq.~(\ref{eq:naiveexample1}). In this normalization, the standard kinetic term is rendered canonical, even though the mixed-derivative term is expected to be the dominant operator that carries time derivatives in the UV.  This does not seem to be a sensible choice of normalization.

The results indeed changes if we choose the normalization in (\ref{eq:mixedlifshitz}) such that $\beta=1$, while still requiring the UV dominant operators to have the same scaling rule. Since the latter condition again imposes $m=z-1$, we now have $[\alpha]=[k]^2$ and $[\gamma]=[k]^0$, leading to the Lagrangian
\begin{equation}
{\cal L}_{2} =M^2 \dot{\phi}^2 -\dot{\phi}\triangle\dot{\phi}-\lambda\,\phi (-\triangle)^z \phi\,.
\label{eq:naiveexample2}
\end{equation}
For this example, the momentum dimension of the scalar field is
\begin{equation}
[\phi] = [k]^{(D-m-2)/2} \,,
\end{equation}
i.e. it is dimensionless for $z=m+1= D-1$, leading to the coefficients of the self-interaction terms to have $[g_n]=[k]^{2(D-1)}$. In $3+1$ dimensions, this corresponds to {\em relativistic} scaling and 4th order gradient terms.

This second example seems to suggest that the mixed derivative term actually improves the UV behavior of the theory. However, the relativistic scaling implies that operators with 4 time derivatives come at the same order as the mixed derivative operator or operators with 4 spatial gradients. With this scaling there is no justification for not including 4th order time derivatives in the action. As is well known, though, including such operators would lead to extra degrees of freedom and potential loss of unitarity. 

\subsection{Superficial degree of divergence}
\label{subsec:superficial}
The existence of two drastically different results for the same theory illustrates that the na\"ive counting method is highly dependent on the choice of scaling and normalization, and can therefore be confusing. Though it does seem straightforward that canonically normalizing the usual kinetic term is not the way to go,
in order to remove any ambiguity we calculate the superficial degree of divergence, in the fashion of Refs.~\cite{Visser:2009fg, Visser:2009ys}. This method allows us to identify the cut-off dependence of the diagrams without relying on the dimensional arguments. 

For the Lagrangian in Eq.(\ref{eq:mixedlifshitz}), the dimensions of the coupling constant are related through
\begin{equation}
[\alpha]\,[k]^{2m} = [\beta]\,[k]^{2m+2}=[\gamma] \,[k]^{2\,z}\,,
\end{equation}
which allows us to rewrite (\ref{eq:mixedlifshitz}) as
\begin{equation}
{\cal L} = \beta\left[\lambda\,M^2\dot{\phi}^2-\dot{\phi}\triangle\dot{\phi} - M^{2(m-z+1)}\,\phi (-\triangle)^z\phi\right]\,.
\end{equation}
Using the equation of motion for the Lifshitz scalar,
\begin{equation}
\beta\,\left[-\lambda\,M^2\,\ddot{\phi}+\triangle\ddot{\phi}-M^{2(m-z+1)}(-\triangle)^z\phi\right]=0\,,
\end{equation}
the Green's function in the UV, i.e. $k \gg \sqrt{\lambda}M$, can be immediately calculated as
\begin{equation}
G_{\omega, k}= \frac{1}{k^2\beta\,[\omega^2-M^{2(m-z+1)}k^{2(z-1)}]}\,.
\label{eq:lifshitzgreen}
\end{equation}
Thus, the dependence of each internal line on the momentum cut-off $\Lambda_k$ is 
\begin{equation}
G_{\omega,k} \to \beta^{-1}M^{-2(m-z+1)}\Lambda_k^{-2z}\,.
\end{equation}

For the loop integrals, we need to impose a different cut-off $\Lambda_\omega$ for the energy. The dependence of the latter on the momentum cut-off can be inferred from the poles of the propagator, giving $\Lambda_\omega= M^{m-z+1}\Lambda_k^{z-1}$. Thus the contribution from each loop in a diagram is 
\begin{equation}
\int d\omega d^Dk \to \Lambda_\omega\,\Lambda_k^D = M^{m-z+1}\Lambda_k^{z+D-1}\,.
\end{equation}

We first consider non-derivative interactions, where the vertices do not contribute to the cut-off dependence. Thus, for a diagram with $I$ internal lines and $L$ loops, the dependence on the momentum cut-off is
\begin{equation}
\beta^{-I} \,M^{(m-z+1)(L-2\,I)}\,\Lambda_k^{L(D+z-1)-2\,I\,z}\,,
\label{eq:mixed-feynmann}
\end{equation}
giving the superficial degree of divergence
\begin{equation}
\delta = (D+z-1)L-2\,I\,z = (D-z-1)L-2\,(I-L)z\,.
\end{equation}
Since $L$ loops require at least $L$ internal lines, we obtain
\begin{equation}
\delta \leq (D-z-1) L\,.
\end{equation}
This implies that if $z\geq D-1$, the diagrams are, at most, logarithmically divergent. For $D=3$, the mixed-derivative theory with relativistic scaling and relativistic dispersion relations is power-counting renormalizable with gradient terms $z\geq 2$. The propagator (\ref{eq:lifshitzgreen}) now contains an overall factor of $k^{-2}$ which ameliorates the UV behavior, alleviating the need for more than $4$ gradients in the action. 

As already mentioned in the previous section, the relativistic scaling is worrisome, as it implies that 4th order time derivative operators are not higher order and should be taken into consideration. Their presence would compromise unitarity without changing the renormalizability properties.
This situation is reminiscent of the renormalization of higher derivative gravity \cite{Stelle:1976gc}. There the dispersion relation is also relativistic and the presence of the higher order derivatives (and the extra degrees of freedom)  improves the UV behavior but breaks the unitarity \cite{Stelle:1976gc}.

The superficial degree of divergence also exposes the limitations of the dimensional counting. In the latter, each momentum dimension is implicitly assumed to contribute one power of the momentum cut-off. However, this assumption is not correct if coefficients of the relevant terms are dimensionful. The dimensional counting can be trusted only in a setup in which $\beta$ and $M$ drop out of the amplitudes; this corresponds to the normalization $\beta=1$ and choice of units with $m=z-1$, which is the second example studied in Sec.\ref{subsec:naivecounting}. This result further demonstrates that the mixed derivative terms cannot be interpreted as deformations of the canonical Lifshitz scalar.

We can further extend the analogy with the Lifshitz scalar to mimic derivative self-interactions of the graviton. Following Ref.~\cite{Visser:2009ys}, we consider the action 
\begin{equation}
{\cal L} = - \dot{\phi}\triangle \dot{\phi} +P(\nabla^{2\,z},\phi)\,,
\end{equation}
where $P(\nabla^{2\,z},\phi)$ is an infinite order polynomial for the field, with up to $2z$ derivatives.
For the free field, i.e. at the quadratic level, the action contains spatial derivative terms up to $\phi\triangle^{z}\phi$, so the propagator in the UV is still given by Eq.(\ref{eq:lifshitzgreen}) with $\beta=1$ and $m=z-1$. The major difference to the previous case comes from the vertices, which can bring at most $2z$ powers of momentum. Thus, the superficial degree of divergence for the diagram with $V$ vertices satisfies
\begin{equation}
\delta  \leq 
(D-z-1)L-2\,(I-L-V)z\,,
\end{equation}
which can be simplified using the topological identity $V + L-I = 1$ to give
\begin{equation}
\delta  \leq  (D-z-1)L+2\,z\,.
\end{equation}
As long as $z\geq D-1$, we have $\delta \leq 2\,z$ where the superficial degree of divergence is bounded from above by the canonical dimension of the operators explicitly included in the bare action. This is an indication of power-counting renormalizability.

\section{Discussion}
\label{sec:discussion}

Ho\v{r}ava gravity has an extra scalar propagating degree of freedom with respect to general relativity. Additionally, usual spin-2 graviton which both theories propagate, has different behavior in Ho\v rava gravity due to the presence of terms with higher-order spatial derivatives in the action.  In contrast, the gauge vector modes do not get any contribution from these higher-derivative terms, thus their propagators are identical to the ones in GR. As a result, as it has been shown in Ref.~\cite{Pospelov:2010mp}, Lorentz violations in the Standard Model sector have quadratic sensitivity to the cut-off stemming from the gauge loops. Supplementing the action with mixed-derivative terms --- terms that contain both temporal and spatial derivatives --- has been suggested as a potential way to regulate these divergences.

We have considered here the most general action of non-projectable Ho\v{r}ava gravity, extended with terms containing two time derivatives and two spatial ones. We have carried out a full perturbative analysis. which revealed that the mixed derivative terms can drastically change the behavior of the propagators. The   
dispersion relations generically become fourth order in the UV, {\em i.e.} $\omega^2 \sim k^4$. This could compromise power-counting renormalizability, which required 6th order dispersion relations in the standard theory. However, we also find that a tuning of the coefficients of the mixed-derivative terms that reinstates the sixth order dispersion relations does exist.

A difficulty one encounters is that renormalizability arguments in standard Ho\v rava gravity are based on anisotropic scaling and on the analogy with the Lifshitz scalar. The mixed-derivative terms do not seem to straightforwardly fit in this logic, and one might rightfully question whether 6th order dispersion relations are really necessary. In order to explore this issue further and avoid the complications that one has to face when dealing with a theory with multiple degrees of freedom, we have considered the Lifshitz scalar itself, extended by adding mixed-derivative terms. We have shown that the mixed-derivative terms actually appear to improve the UV behavior and the theory can be renormalizable even with 4th order dispersion relations. However, this comes at a high price: the scaling between space and time is actually relativistic and terms with 4th order time derivatives appear to come at the same order as those included in the action. Hence, one expects that this theory will cease to be unitary once quantum corrections are taken into account. 

Therefore, to the extent that one can transfer the intuition coming from the Lifshitz scalar to Ho\v rava gravity, tuning the coefficients of the mixed-derivative terms so as to have 6th order dispersion relations and anisotropic scaling seems preferable. 
Note that such a tuning does not obstruct the effect of the mixed-derivative terms on the gauge modes. This is particularly important in order to suppress the Lorentz violations in the matter sector (it is the motivation for adding mixed-derivative terms in the first place).
However, the pertinent question is if such a tuning could be technically natural.

Our whole analysis is based on linearized theory (as is power-counting renormalizability in the first place). The tuning appears technically natural in linearized theory but our approach cannot address radiative stability beyond the linear level. More work in this direction is needed in order to conclude if adding mixed-derivative terms in Ho\v rava gravity is a viable way to cure the quadratic divergencies related to the vector mode found in Ref.~\cite{Pospelov:2010mp}.

\begin{acknowledgments}
We are grateful to Jorma Louko, Maxim Pospelov and Matt Visser for a critical reading of the manuscript and helpful comments.
The research leading to these results has received funding from the European Research
Council under the European Union's Seventh Framework Programme
(FP7/2007-2013) / ERC Grant Agreement n. 306425 ``Challenging General
Relativity''.  
\end{acknowledgments}

\end{document}